\documentclass[seceq]{ptptex}

\usepackage{graphicx}

\newcommand{\vi}[1]{\mbox{\boldmath $#1$}}
\newcommand{\vis}[1]{\mbox{\boldmath ${\scriptstyle #1}$}}

\title{Green's function method for strength function\\ 
in three-body continuum%
}

\author{
Yasuyuki \textsc{Suzuki}$^{1,}$\footnote{E-mail: suzuki@nt.sc.niigata-u.ac.jp},
 Wataru \textsc{Horiuchi}$^{2,}$\footnote{E-mail: horiuchi@nt.sc.niigata-u.ac.jp},
and Daniel \textsc{Baye$^{3,}$\footnote{E-mail: dbaye@ulb.ac.be}}
}

\inst{$^1$Department of Physics, Faculty of Science 
and Graduate School of Science and Technology, Niigata University, Niigata
950-2181, Japan\\
$^2$Graduate School of Science and Technology, Niigata
University, Niigata 950-2181, Japan\\
$^3$Physique Nucl\'{e}aire Th\'{e}orique et Physique
Math\'{e}matique, C.P. 229, Universit\'{e} Libre de
Bruxelles, B 1050, Belgium
}

\abst{Practical methods to compute dipole strengths for a three-body system 
by using a discretized continuum are analyzed. 
New techniques involving Green's function are developed, either by 
correcting the tail of the approximate wave function in a direct 
calculation of the strength function or by using a solution 
of a driven Schr\"odinger equation in a summed expression of the strength. 
They are compared with the complex scaling method and the Lorentz 
integral transform, also making use of a discretized continuum. 
Numerical tests are performed with a hyperscalar three-body 
potential in the hyperspherical-harmonics formalism. 
They show that the Lorentz integral transform method is less practical 
than the other methods because of a difficult inverse transform. 
These other methods provide in general comparable accuracies. 
}

\PTPindex{205}

\begin{document}

\maketitle

\section{Introduction}

{\it Ab initio} calculations of the strength function of a 
quantum system for a perturbation play a very important 
role to understand its resonances and continuum states. 
It is meant by {\it ab initio} that the calculation requires 
only a Hamiltonian and a physical condition of the system but is not 
based on specific model assumptions like a mean field model.  
This approach has attracted increasing interest 
with the advent of halo 
nuclei which are typical examples of a weakly bound system. 
These nuclei have only one or few bound states and most 
phenomena involving them are properly understood by paying 
due attention to the role of the continuum and the resonances. 

Three-body continuum states are needed to calculate  
the energy spectra of the particles in the breakup of 
two-neutron halo nuclei~\cite{danilin,danilin06} as well as 
the three-body partial decay widths of nuclei~\cite{garrido}.
Though the construction of the three-body continuum states is 
elegantly formulated in hyperspherical harmonics (HH) 
methods~\cite{danilin}, 
numerical computation generally involves heavy tasks and suffers 
from the problem of slow convergence~\cite{bk,barletta}.  

Several methods have been proposed to resolve this problem 
within the HH formalism~\cite{nielsen,desc06}. 
Another class of powerful approaches is to avoid a direct 
construction of the continuum 
but to utilize continuum-discretized states (CDS)  
as in the complex scaling method (CSM)~\cite{aoyama} 
and the Lorentz integral transform (LIT)~\cite{efros} method. 
Some of these methods are in fact applied to studying the breakup 
cross section and 
the electric dipole strength of the 
two-neutron halo nucleus $^6$He~\cite{baye,myo,egami}. 

The purpose of the present paper is twofold: The first is to 
propose a new method of generating three-body 
continuum in the HH formalism. The second is to assess the 
efficiency of this approach in a simple solvable problem 
by calculating the electric dipole strength 
function of a three-body system and comparing it to those of 
the CSM and LIT method. The basic vehicle of the present approach is 
a Green's function, the power of which is widely recognized 
in many problems and very recently proved in the phase-shift 
calculation of scattering between nuclei~\cite{ps.cal}. 
Ill tail behavior of the wave function 
obtained in a calculation using square-integrable basis functions 
is appropriately corrected with use of the Green's function. 

In Sect.~\ref{sec.str.fn} we briefly explain how 
the strength function is calculated not only in 
the CSM and LIT method but also in an approach based on the 
driven equation of motion (DEM). In Sect.~\ref{sec.three-body} 
we present a Green's function approach to solve 
the three-body equation in the HH formalism. In solving 
both standard and driven equations of motion square-integrable bases 
are used in advantage owing to the Green's function. 
In Sect.~\ref{sec.example} we treat a specific example of a 
three-body Hamiltonian which allows numerically 
exact calculations of the electric 
dipole strength function and thus enables us to 
compare to those obtained in various methods. 
Conclusion is drawn in Sect.~\ref{sec.conclusion}. 
Appendix A presents 
a simple method of solving an inhomogeneous equation with 
an outgoing-wave boundary condition. 
The subtleties in the expansion of the complex scaled resolvent, 
which is at the heart of the CSM, 
are noted in Appendix B for the sake of completeness. 
The analytic form of the electric 
dipole strength 
for the three-particle continuum is discussed in Appendix C.

\section{Methods of strength function calculation}
\label{sec.str.fn}

The physics we treat here involves the process that a system 
in its ground state is excited 
to three-body continuum states by a perturbation. 
Let $H$ be the Hamiltonian of the system and $W$ be the 
perturbation. We want 
to calculate the strength or response function given 
in lowest order perturbation theory as 
\begin{equation}
S(E)=\sum_{\nu} |\langle \Psi_{\nu}|W|\Psi_0 \rangle|^2 \delta(E_{\nu}-E),
\label{str.fn}
\end{equation}
where $\Psi_0$ is the ground state with energy $E_0$, and 
$\Psi_{\nu}$ is the excited state with energy $E_{\nu}$: 
$H\Psi_0=E_0\Psi_0$ and $H\Psi_{\nu}=E_{\nu}\Psi_{\nu}$. 
The energy 
is measured from some standard value, e.g., a particle-decay 
threshold energy. 
When the excited state is in the continuum, the
label $\nu$ is continuous and the sum is replaced by an
integration. In case where the energy $E_{\nu}$
is degenerate, the sum also runs over the degenerate 
states. The continuum eigenstate 
$\Psi_{\nu}$ is normalized on an energy scale, i.e.,  
$\langle \Psi_{\nu} | \Psi_{\nu'} \rangle =\delta(E_{\nu}-E_{\nu'})$.

It is convenient to derive an expression in which 
the summation in Eq.~(\ref{str.fn}) becomes implicit. 
This is possible if Eq.~(\ref{str.fn}) is rewritten as 
\begin{equation}
S(E)=\langle \Psi_0|W^{\dagger}\delta(H-E)W|\Psi_0 \rangle,
\label{str.fn0}
\end{equation}
which can be further rewritten as 
\begin{equation}
S(E)=-\frac{1}{\pi}{\rm Im}
\langle \Psi_0|W^{\dagger}{\cal G}(E+i\epsilon)W|
\Psi_0 \rangle.
\label{str.fn1}
\end{equation}
Here 
\begin{equation}
{\cal G}(E)=\frac{1}{E-H}
\end{equation}
is the resolvent, and use is made of 
the identity 
${\cal G}(E+i\epsilon)={\rm P}{\cal G}(E)-i\pi \delta(E-H)$, where 
P stands for taking Cauchy's principal value of the integral. 
The outgoing-wave boundary condition is chosen by including $i\epsilon$ 
in the resolvent, where $\epsilon$ is a positive
infinitesimal value. It is noted that formulas for sum rule 
values, 
$\int E^n S(E)dE\,$\\$ (n=0,1,...)$, are easily derived with use of 
Eq.~(\ref{str.fn0}). 

The calculation of the strength function 
using Eq.~(\ref{str.fn1}) 
was performed in Ref.~\citen{shlomo} to include the continuum 
effects of particle-hole excitations in the random-phase 
approximation, and recently applied in the 
Hartree-Fock-Bogoliubov formalism~\cite{matsuo,khan}. 
A new approach as well as some existing methods to calculate $S(E)$ 
for more general cases are explained in the following subsections.

\subsection{Direct calculation with continuum discretized states}

The first and most obvious approach is to calculate expression (\ref{str.fn}) 
directly. 
Because of the selection rules, only a limited number of wave functions must 
be computed at each energy. 
We shall use this technique to provide a reference calculation in 
Sect.~\ref{sec.example} by computing the 
bound-state and scattering wave 
functions numerically with finite differences. 

However, our goal here is to show that the direct approach can be performed 
with square-integrable basis functions. 
Therefore, we can extend the Green's function technique developed 
in Ref.~\citen{ps.cal} to derive approximate scattering states directly 
usable in Eq.~(\ref{str.fn}).
This direct method will be referred to as the continuum-discretized 
approximation (CDA).

\subsection{Driven equation of motion method}

Let $\Psi$ denote 
\begin{equation}
\Psi={\cal G}(E+i\epsilon)W\Psi_0.
\end{equation}
The wave function $\Psi$ is a solution of the driven Schr\"{o}dinger equation
\begin{equation}
(H-E)\Psi=-W\Psi_0
\label{driven.eq}
\end{equation}
with the outgoing-wave boundary condition in the 
asymptotic region. Thus we obtain 
\begin{equation}
S(E)=\frac{1}{\pi}{\rm Im}\langle \Psi |W|\Psi_0 \rangle
=-\frac{1}{\pi}{\rm Im}\langle \Psi |H-E|\Psi \rangle.
\label{str.fn2}
\end{equation}
Solving Eq.~(\ref{driven.eq}) with the outgoing-wave boundary condition 
is the main task in the DEM method. 
This approach is used to describe the double photoionization 
of a two-electron atom~\cite{pont,vanroose}. 
To cope with the boundary 
condition of the outgoing wave, the exterior complex 
scaling is often employed~\cite{mcmurdy}. Instead of using the exterior 
complex scaling, we will 
present in Sect.~\ref{formalism} and Appendix A a method of 
solving Eq.~(\ref{driven.eq}) with use of 
the Green's function. 

\subsection{Complex scaling method}
\label{csm}
In the CSM 
the strength function is evaluated using the expression
\begin{equation}
S(E)=-\frac{1}{\pi}{\rm Im}\langle \Psi_0|W^{\dagger}
U^{-1}(\theta)R(\theta)U(\theta)W|\Psi_0 \rangle,
\label{dbdecsm}
\end{equation}
which is readily obtained from Eq.~(\ref{str.fn1}), 
where $U(\theta)$ is an unbounded operator which transforms 
all the coordinates 
used to specify the system as ${\vi x}\to {\rm e}^{i\theta}{\vi x}$, 
and $U^{-1}(\theta)=(U(\theta))^{-1}$. 
Here $R(\theta)$ is 
the complex scaled resolvent $R(\theta)=U(\theta){\cal G}(E+i\epsilon)
U^{-1}(\theta)
=1/(E-H(\theta)+i\epsilon)$ with $H(\theta)
=U(\theta)HU^{-1}(\theta)$.

Since any continuum eigenstate of $H$ that has 
an outgoing wave in the asymptotic region is transformed into
a function which damps at large distances 
within a suitable choice of $\theta$, 
it is possible to expand the complex scaled eigenfunction over a set of
linearly independent square-integrable basis functions 
$\Phi_i(\vi x)$~\cite{lowdin,kruppa}:
\begin{equation}
H(\theta)\Psi^{\lambda}(\theta)
=E^{\lambda}(\theta)\Psi^{\lambda}(\theta),
\label{diag.csm}
\end{equation}
with
\begin{equation}
\Psi^{\lambda}(\theta)=\sum_{i}C^{\lambda}_i(\theta)\Phi_i(\vi x).
\end{equation}
Here $\lambda$ is a label to characterize the eigenfunction. 
Substituting this expansion into Eq.~(\ref{diag.csm}) enables 
one to obtain $E^{\lambda}(\theta)$ and $C_i^{\lambda}(\theta)$.  
The eigenfunction expansion of $R(\theta)$ leads to the 
expression 
\begin{equation}
S(E)
=-\frac{1}{\pi}\sum_{ \lambda}{\rm Im}
\frac{{\widetilde{\cal D}}^{\lambda}(\theta){\cal
D}^{\lambda}(\theta)}{E-E^{\lambda}(\theta)+i\epsilon},
\label{str.csm}
\end{equation}
with
\begin{equation}
{\cal D}^{\lambda}(\theta)=
\langle (\Psi^{\lambda}(\theta))^*|W(\theta)|U(\theta)\Psi_0 \rangle,
\ \ \ \ \ 
\widetilde{{\cal D}}^{\lambda}(\theta)=
\langle (U(\theta)\Psi_0)^*
|W^{\dagger}(\theta)|\Psi^{\lambda}(\theta)\rangle,
\label{def.DD}
\end{equation}
where $W(\theta)=U(\theta)WU^{-1}(\theta)$, 
$W^{\dagger}(\theta)=
U(\theta)W^{\dagger}U^{-1}(\theta)$, 
and $U(\theta)\Psi_0$ is the solution of Eq.~(\ref{diag.csm})
corresponding to the ground state, the eigenvalue 
$E^{\lambda}(\theta)$ of which should be equal to $E_0$ in principle. 
See Appendix B for the derivation of Eqs.~(\ref{str.csm}) 
and (\ref{def.DD}). 
The accuracy of $S(E)$ calculated with Eq.~(\ref{str.csm}) is tested 
by observing its stability against $\theta$. 
See Ref.~\citen{aoyama} for the details, performances and references 
of the CSM.

The contribution of a bound excited state to $S(E)$ is not 
calculated from Eq.~(\ref{str.csm}) because its energy 
$E^{\lambda}(\theta)$ becomes real in principle independently of $\theta$, but 
is separately calculated from the original 
equation~(\ref{str.fn}). 

\subsection{Lorentz integral transform method}
\label{sec.litm}

Let us define the Lorentz transform of the strength function
\begin{equation}
{\cal L}(z)=\int_{E_{\rm min}}^{\infty}\frac{S(E)}{(E-z)(E-z^*)}dE
=\int_{E_{\rm min}}^{\infty}\frac{S(E)}{(E-E_R)^2+E_I^2}dE,
\label{L-S.relation}
\end{equation}
where $z=E_R+iE_I$ is a complex energy and $E_{\rm min}$ is a minimum
energy from which $S(E)$ begins to have strength. 
Substitution of Eq.~(\ref{str.fn}) leads to
\begin{eqnarray}
{\cal L}(z)=\sum_{\nu}\frac{|\langle \Psi_{\nu}|W|\Psi_0
 \rangle|^2}{(E_{\nu}-z)(E_{\nu}-z^*)}
=\langle \Psi_0 |W^{\dagger}{\cal G}(z^*){\cal G}(z)W|
\Psi_0\rangle.
\end{eqnarray}
Thus ${\cal L}(z)$ reduces to the overlap, 
${\cal L}(z)=\langle \Psi(z)|\Psi(z) \rangle$,
of $\Psi(z)={\cal G}(z)W\Psi_0$. The function $\Psi(z)$ 
satisfies the following system of equations 
\begin{equation}
(H-z)\Psi(z)=-W\Psi_0.
\label{litm}
\end{equation}
This equation has the same structure as 
Eq.~(\ref{driven.eq}) of the DEM method, but their solutions have quite 
different asymptotic behavior. 
Since ${\cal L}(z)$ takes a finite value in so far as $z$ is complex, 
$\Psi(z)$ has a finite norm, that is, it damps at large distances. 
Therefore $\Psi(z)$ can be expanded in terms of 
square-integrable basis functions $\Phi_i({\vi x})$ as 
\begin{equation}
\Psi(z)=\sum_i C_i(z)\Phi_i({\vi x}),
\end{equation}
and the coefficients $C_i(z)$ are determined from the 
equation 
\begin{equation}
\sum_j \langle \Phi_i|H-z|\Phi_j\rangle C_j(z)=-\langle
 \Phi_i|W|\Psi_0 \rangle.
\end{equation}

An accurate calculation of ${\cal L}(z)$ is usually not a major 
problem, but determining 
$S(E)$ is the main task in the LIT method 
because it requires the inversion of 
Eq.~(\ref{L-S.relation}). To perform the inversion, 
${\cal L}(z)$ values are calculated for a number of $E_R$ 
values in a very wide interval for some chosen $E_I$ value. 
This is necessary in order not to miss 
the sum rule of $S(E)$, which is related to the integral 
of ${\cal L}(z)$ by 
\begin{equation}
\int_{E_{\rm min}}^{\infty}S(E)dE=\frac{E_I}{\pi}\int_{-\infty}^{\infty}
 {\cal L}(z) dE_R. 
\end{equation}
One assumes a plausible 
form of $S(E)$ which contains some parameters, and then determines 
those parameters 
so as to reproduce the ${\cal L}(z)$ data as accurately as 
possible. One has to make sure that $S(E)$ 
determined in this way is stable independently 
of the choice of $E_I$. See a review article~\cite{efros} 
for the LIT method and the relevant literature.

\section{Theory for three-body continuum}
\label{sec.three-body}

\subsection{Hyperspherical harmonics method}

We use a three-body system to compare the electric dipole 
strength functions 
calculated in various methods. To make this article self-contained, 
we give some basic formulas which are needed to formulate the dynamics 
of three particles in the HH method. 
See Refs.~\citen{danilin,nielsen} for details.

Let $A_1m,A_2m,A_3m$ and $Z_1e, Z_2e, Z_3e$ be the masses 
and charges of the three particles, where $m$ 
is a nucleon mass and $e$ is the electron charge magnitude. 
For the sake of simplicity we assume that two of the 
three particles are neutral, as in $p+n+n$ and $\alpha+n+n$, so 
that no Coulomb potential acts among the particles. 

Let $H=T+V$ be the Hamiltonian of the three-body system
with 
\begin{equation}
T=T_1+T_2+T_3-T_{\rm cm},\ \ \ \ \ V= V_{12}+V_{23}+V_{31}+V_{123},
\label{T-V.def}
\end{equation}
where $T_i$ is the kinetic energy of particle $i$, and the kinetic
energy of the center of mass motion, $T_{\rm cm}$,  
is subtracted in the Hamiltonian. 
Here $V_{ij}$ is the nuclear interaction acting 
between $ij$ pair and $V_{123}$ is a nuclear three-body force.  

The Jacobi relative coordinates ${\vi x}_1$ and ${\vi x}_2$ are 
defined as
\begin{equation}
{\vi x}_1=\sqrt{A_{1,2}}({\vi r}_1-{\vi r}_2),\ \ \ \ \ 
{\vi x}_2=\sqrt{A_{12,3}}\left(\frac{A_1{\vi
r}_1+A_2{\vi r}_2}{A_1+A_2}-{\vi r}_3\right),
\end{equation}
where ${\vi r}_i$ is the position vector of particle $i$, and 
$A_{i,j}={A_iA_j}/(A_i+A_j)$ and  
$A_{ij,k}=(A_i+A_j)A_k/(A_i+A_j+A_k)$
are the reduced mass factors. 
Let ${\vi x}_3$ denote the center of mass coordinate, 
${\vi x}_3=(1/A)\sum_{i=1}^3A_i{\vi r}_i$, where $A=A_1+A_2+A_3$.

The electric dipole operator $W_{\mu}$ of the system is given by
\begin{equation}
W_{\mu}=\sqrt{\frac{3}{4\pi}}
\sum_{i=1}^3 Z_ie({\vi r}_i-{\vi x}_3)_{\mu}
=\sqrt{\frac{3}{4\pi}}(d_1{{\vi x}_1}_{\mu}+d_2{{\vi x}_2}_{\mu}),
\label{E1.op}
\end{equation}
with
\begin{eqnarray}
d_1&=&\frac{e}{\sqrt{A_1A_2(A_1+A_2)}}(Z_1A_2-Z_2A_1),
\nonumber \\
d_2&=&\frac{e}{\sqrt{(A_1+A_2)A_3A}}((Z_1+Z_2)A_3-Z_3(A_1+A_2)).
\end{eqnarray}
Note that 
the quantity, $d_1^2+d_2^2$,  reduces to 
$d_1^2+d_2^2=e^2\left\{\sum_i(Z_i^2/A_i)-(\sum_i Z_i)^2/A\right\}$.

The HH method is convenient to study 
the three-particle dynamics. 
The hyperradius and hyperangle coordinates, $\rho$ and $\alpha$, 
are introduced as 
\begin{equation}
\rho=\sqrt{x_1^2+x_2^2},\ \ \ \ \ 
\alpha=\tan^{-1}(x_1/x_2).
\end{equation}
Note that $x_1=\rho \sin \alpha$, $x_2=
\rho \cos \alpha$ with 0 $\leq$ $\alpha$ $\leq$ $\pi/2$. 
The five angle coordinates, $\alpha$ as well as 
$\theta_1, \phi_1, \theta_2, \phi_2$ of $\hat{{\vi x}}_1$, 
$\hat{{\vi x}}_2$, are denoted by $\Omega$ collectively. 
The volume element reads 
\begin{equation}
d{\vi r}_1d{\vi r}_2d{\vi r}_3={\cal J}^2\rho^5d\rho d{\Omega}d{\vi x}_3
\end{equation}
with
\begin{equation}
{\cal J}=\left(\frac{A}{A_1A_2A_3}\right)^{3/4},
\label{jacobian}
\end{equation}
where $d{\vi x}_1d{\vi x}_2=\rho^5d\rho d{\Omega}$, and 
$d\Omega=\sin^2\alpha \cos^2 \alpha \, 
d\alpha \, d{\hat{\vi x}_1}\,d{\hat{\vi x}_2}$. 
It is noted that the squared hyperradius $\rho^2$ is related to 
the mean squared radius, $\rho^2=\sum_{i=1}^3A_i({\vi r}_i-{\vi
x}_3)^2$, or to the sum of the squared relative distances between the
particles, $\rho^2=(1/A)\sum_{i<j}A_iA_j({\vi r}_i-{\vi r}_j)^2$.

The kinetic energy operator $T$ in Eq.~(\ref{T-V.def}) reads 
\begin{equation}
T=-\frac{\hbar^2}{2m}\left(\frac{\partial^2}{\partial {\vi x}_1^2}
+\frac{\partial^2}{\partial {\vi x}_2^2}\right)
=-\frac{\hbar^2}{2m}\left(\frac{\partial^2}{\partial
		       \rho^2}+\frac{5}{\rho}
\frac{\partial}{\partial \rho}-\frac{1}{\rho^2}{\cal K}^2
\right),
\end{equation}
with the hypermomentum operator 
\begin{equation}
{\cal K}^2=-\frac{\partial^2}{\partial \alpha^2}-4\cot 2\alpha 
\frac{\partial}{\partial \alpha}+\frac{1}{\sin^2 \alpha}
{\vi \ell}_1^2+\frac{1}{\cos^2 \alpha}{\vi \ell}_2^2,
\end{equation}
where ${\vi \ell}_1, {\vi \ell}_2$ are the angular momenta 
corresponding to the coordinates ${\vi x}_1, {\vi x}_2$, 
respectively. 

The normalized eigenfunction of ${\cal K}^2$, called the HH,  
with eigenvalue $K(K+4)$ is given by 
\begin{equation}
{\cal F}_{KLM_L}^{\ell_1 \ell_2}(\Omega)=\phi_{K}^{\ell_1
 \ell_2}(\alpha)[Y_{\ell_1}(\hat{\vi x}_1)Y_{\ell_2}(\hat{\vi
 x}_2)]_{LM_L},
\label{eigfn.K}
\end{equation}
where $K$ is an integer called the hypermomentum, and 
\begin{equation}
\phi_{K}^{\ell_1 \ell_2}(\alpha)={\cal N}_K^{\ell_1 \ell_2}
\sin^{\ell_1}\alpha \cos^{\ell_2}\alpha \, 
G_n(\ell_1+\ell_2+2,\ell_1+\textstyle{\frac{3}{2}}; \sin^2\alpha), 
\end{equation}
where $n$ is an integer given by 
$n$=$(K-\ell_1-\ell_2)/2$ and the Jacobi polynomial $G_n$ is 
expressed in terms of the Gauss hypergeometric series as 
$G_n(\ell_1+\ell_2+2,\ell_1+\textstyle{\frac{3}{2}};z^2)=F(-n,\ell_1+\ell_2+n+2,
\ell_1+\textstyle{\frac{3}{2}};z^2)$ and
\begin{equation}
{\cal N}_K^{\ell_1 \ell_2}=\sqrt{\frac{2(K+2)\Gamma(\ell_1+\ell_2+n+2)\Gamma(\ell_1+n+
\textstyle{\frac{3}{2}})}{n!\Gamma(\ell_2+n+\textstyle{\frac{3}{2}})
[\Gamma(\ell_1+\textstyle{\frac{3}{2}})]^2}}.
\end{equation}
The function 
${\cal F}_{KLM_L}^{\ell_1 \ell_2}$ is usually denoted 
${\cal Y}_{KLM_L}^{\ell_1 \ell_2}$ in literatures. 
Using the orthogonality of $G_n$ leads to the following 
orthonormality relation 
\begin{equation}
\int ({\cal F}_{KLM_L}^{\ell_1 \ell_2}(\Omega))^*
{\cal F}_{K'L'M_L'}^{\ell'_1 \ell'_2}(\Omega)d\Omega
=\delta_{\ell_1, \ell'_1}\delta_{\ell_2, \ell'_2}
\delta_{K, K'}\delta_{L, L'}\delta_{M_L, M_L'}.
\label{ortho}
\end{equation}
We also note the completeness relation
\begin{equation}
\delta(\Omega-\Omega')=\sum_{\ell_1 \ell_2 KLM_L}
{\cal F}_{KLM_L}^{\ell_1 \ell_2}(\Omega)
({\cal F}_{KLM_L}^{\ell_1 \ell_2}(\Omega'))^*.
\label{complete}
\end{equation}

\subsection{Solving with Green's function}
\label{formalism}

Ignoring the spin and isospin degrees of freedom of the particles, 
we focus on the spatial part of the wave function. The antisymmetry 
requirement on the wave function is also ignored. 
The inclusion of these  
causes no problem and will be detailed in a separate paper.  

We want to obtain the continuum state $\Psi$ for 
the equation of motion of type
\begin{equation}
(H-E)\Psi=\Phi
\label{sch.eq}
\end{equation}
for a given $\Phi$. 
Here $E$ is a given energy from the three-particle threshold.  
There are two cases of our interest. In the first case, 
$\Phi=0$ and $\Psi$ is the 
continuum state to be used for $\Psi_{\nu}$ in 
Eq.~(\ref{str.fn}). The second case is concerned with the 
DEM method, that is $\Phi=-W_{\mu}\Psi_0$,  
and we want to find such $\Psi$ that has the outgoing wave 
in the asymptotic region, as required in Eq.~(\ref{driven.eq}). 
In both cases we seek 
$\Psi$ with the angular momentum $L$ and 
its projection $M_L$ in an expansion in terms of 
the channel wave functions 
${\cal F}_{cLM_L}={\cal F}^{\ell_1 \ell_2}_{KLM_L}(\Omega)$
\begin{equation}
\Psi= \rho^{-5/2}\sum_c f_c(\rho){\cal F}_{cLM_L},
\label{exp.HH}
\end{equation} 
where $c=(\ell_1 \ell_2 K)$ stands for a set of channel labels. 
The aim is now to obtain the hyperradial function $f_c(\rho)$. 

Owing to the orthonormality~(\ref{ortho}), 
$f_c(\rho)$ is related to $\Psi$ as
\begin{equation}
f_c(\rho)=\rho^{5/2} \langle {\cal F}_{cLM_L}|\Psi\rangle.
\label{spectr.amp}
\end{equation}
Substituting Eq~(\ref{exp.HH}) to Eq.~(\ref{sch.eq}) and 
projecting it to the channel $c$, we obtain 
\begin{equation}
\Big(\frac{d^2}{d\rho^2}
-\frac{(K+\frac{3}{2})(K+\frac{5}{2})}{\rho^2}+k^2\Big)f_{c}(\rho)=
\frac{2m}{\hbar^2}\big\{z_c(\rho)+\zeta_c(\rho)\big\},
\label{eq.f-z}
\end{equation}
with
\begin{equation}
z_c(\rho)=\rho^{5/2}\langle {\cal F}_{cLM_L}|V|\Psi\rangle,
\ \ \ \ \ 
\zeta_c(\rho)=-\rho^{5/2}\langle {\cal F}_{cLM_L}|\Phi \rangle,
\label{zzeta}
\end{equation}
where $k^2=2m E/\hbar^2$. Note that the function $z_c(\rho)$ is unknown. 
By using the expansion  
\begin{equation}
z_c(\rho)=\sum_{c'}\langle {\cal F}_{cLM_L}|V|{\cal F}_{c'LM_L}
\rangle f_{c'}(\rho)=\sum_{c'} V_{cc'}(\rho)f_{c'}(\rho)
\label{z.exp}
\end{equation}
as usually done,
Eq.~(\ref{eq.f-z}) becomes a set of 
coupled equations for $f_c(\rho)$, but we keep the 
form~(\ref{eq.f-z}) in order to develop our approach. 
The function $\zeta_c(\rho)$ is known: It is 
either zero for $\Phi=0$ or 
$\rho^{5/2}\langle {\cal F}_{cLM_L}|W_{\mu}
|\Psi_0\rangle$ for $\Phi=-W_{\mu}\Psi_0$. It is important to 
realize that both of 
$z_c(\rho)$ and $\zeta_c(\rho)$ are finite-ranged, that is,  
they vanish for large $\rho$. This is because $V$ is finite-ranged 
and the ground state wave function $\Psi_0$ is spatially confined. 
In actual cases $z_c(\rho)$ is known to decrease in power law
$\rho^{-n}\, (n\geq 3)$ even for a pairwise short-ranged nuclear 
interaction~\cite{thompson}. Moreover, when the Coulomb potential 
is present, the coupling between the different channels 
persists~\cite{desc06}. These problems will make $z_c(\rho)$ decrease 
very slowly.

Let $v_c(\rho)$ and $h_c(\rho)$ denote, respectively, the regular 
and irregular solutions of the homogeneous equation 
with the right-hand side of 
Eq.~(\ref{eq.f-z}) set to zero. They are given in 
terms of the Bessel functions of the first and second kinds
\begin{equation}
v_c(\rho)=(k\rho)^{1/2}J_{K+2}(k \rho),\ \ \ \ \ 
h_c(\rho)=(k\rho)^{1/2}Y_{K+2}(k \rho).
\end{equation}
They satisfy the Wronskian relation, 
$W(v_c,h_c)\equiv v_c(\rho)h_c'(\rho)-v_c'(\rho)h_c(\rho)$=
$2k/\pi$, where $h_c'(\rho)=(d/d\rho)h_c(\rho)$ etc. Now 
we discuss how to solve Eq.~(\ref{eq.f-z}) for the two cases.\\

\par\noindent (1)\,{\it {CDA case}}

The function $f_c(\rho)$ must be regular at $\rho=0$. 
By noting $\zeta_c(\rho)=0$, the formal solution 
of Eq.~(\ref{eq.f-z}) can be written as 
\begin{equation}
f_c^{\rm GF}(\rho)
=\lambda_cv_c(\rho)+\frac{2m}{\hbar^2}\int_0^{\infty}
G_c(\rho,\rho')z_c(\rho')d\rho',
\label{spectr.gf}
\end{equation}
where $\lambda_c$ is a constant yet 
to be determined. The Green's function $G_c$ is a solution of the 
equation
\begin{equation}
\Big(\frac{d^2}{d\rho^2}-\frac{(K+\frac{3}{2})(K+\frac{5}{2})}{\rho^2}
+k^2\Big)G_c(\rho, \rho')
=\delta(\rho-\rho'),
\label{greenf}
\end{equation}
and it is given by
\begin{equation}
G_c(\rho,\rho')=\frac{\pi }{2k}  v_c(\rho_<)h_c(\rho_>),
\label{ggfn}
\end{equation}
where $\rho_< \,(\rho_>)$ is the lesser (greater) of $\rho$ and 
$\rho'$. The asymptotic behavior 
\begin{equation}
f_c^{\rm GF}(\rho)\, \mathop{\longrightarrow}\limits_{\rho \rightarrow \infty} 
\lambda_c \{v_c(\rho) - \tan \delta_c h_c(\rho)\}
\label{asym}
\end{equation}
provides the phase shift $\delta_c$ by 
\begin{equation}
\tan \delta_c=-\frac{\pi m}{\hbar^2 k
 \lambda_c}\int_0^{\infty}v_c(\rho)z_c(\rho)d\rho.
\label{ps.formula}
\end{equation}
The normalization of 
$f_c^{\rm GF}(\rho)$ will be discussed in Sect.~\ref{normalization}.

We have to know $z_c(\rho)$ and determine $\lambda_c$. 
A basic idea for resolving this problem is to 
make use of the CDS following 
Ref.~\citen{ps.cal}. The CDS $\Psi^{\rm CD}$ are 
obtained by diagonalizing the Hamitonian in a 
certain square-integrable basis set which can accurately 
describe the wave function in the interaction region. 
As mentioned above, $z_c(\rho)$ is 
finite-ranged, so that the ill tail behavior of $\Psi^{\rm CD}$ 
causes no problem for evaluating $z_c(\rho)$ 
reliably. 
We only need to require that $\Psi^{\rm CD}$ is 
accurate in the internal region where the potential $V$ is 
effective. Thus we replace $z_c(\rho)$ in Eq.~(\ref{spectr.gf}) by 
$z_c^{\rm CD}(\rho)=\rho^{5/2}\langle {\cal F}_{cLM_L}|V|\Psi^{\rm CD}\rangle$. 
Moreover $f_c(\rho)$ calculated from Eq.~(\ref{spectr.amp}) using 
$\Psi^{\rm CD}$, denoted $f_c^{\rm CD}(\rho)
=\rho^{5/2}\langle {\cal F}_{cLM_L}|\Psi^{\rm CD}\rangle$, 
is expected to 
be already accurate in the internal region. Comparing it with 
$f_c^{\rm GF}(\rho)$ in the internal 
region enables us to determine $\lambda_c$. 
The procedure to determine $\lambda_c$ is 
as follows. Let $[0,a_c]$ be the 
internal region. Choosing $M$ sampling points $(\rho_1,\rho_2,
\ldots,\rho_M) \,(\rho_i \in [0,a_c])$, we determine 
$\lambda_c$ by the least squares fitting
\begin{equation}
{\rm minimize\  over\  \lambda_c}: \sum_{i}
[f_c^{\rm GF}(\rho_i)-f_c^{\rm CD}(\rho_i)]^2.
\end{equation}
In this way we have the continuum state at hand and hence 
can calculate 
the strength function directly. 

As seen above, the energy $E$ of Eq.~(\ref{sch.eq}) cannot be chosen 
arbitrarily in the CDA, but has to be set to the discretized energies 
determined by the diagonalization.\\

\par\noindent (2)\,{\it {DEM case}}

In the DEM method we have to solve Eq.~(\ref{eq.f-z}) 
with $\zeta_c(\rho)\neq 0$. A general solution that is regular at 
$\rho=0$ and has an outgoing wave in the asymptotic region reads 
\begin{equation}
f_c(\rho)=\frac{2m}{\hbar^2}
\int_0^{\infty}G_c^+(\rho,\rho')\big\{z_c(\rho')+\zeta_c(\rho')\big\} d\rho'.
\label{y.formal}
\end{equation}
The Green's function $G_c^+$ satisfying the outgoing-wave 
boundary condition reads 
\begin{equation}
G_c^+(\rho,\rho')=
\frac{\pi }{2ki}  v_c(\rho_<)h_c^+(\rho_>), 
\end{equation}
with
\begin{equation}
h_c^+(\rho)=v_c(\rho)+ih_c(\rho).
\end{equation}

The function $z_c(\rho)$ is again unknown. We note that 
$f_c(\rho)$ and $z_c(\rho)$ are mutually linked by 
Eqs.~(\ref{z.exp}) and (\ref{y.formal}). A method to 
determine $f_c(\rho)$ is as follows. Because 
both $z_c(\rho)$ and $\zeta_c(\rho)$ are finite-ranged, that is,  
negligible for, say $\rho \geq a_c$, it is found that 
$f_c(\rho)$ takes the form for $\rho \geq a_c$
\begin{equation}
f_c(\rho)=A_ch_c^+(\rho),
\label{fcout}
\end{equation}
with a constant $A_c$ 
\begin{equation}
A_c=\frac{\pi}{2ki}\int_0^{a_c} v_c(\rho)\big\{z_c(\rho)+\zeta_c(\rho)\big\}
d\rho.
\end{equation}
Therefore we need to determine $f_c(\rho)$ in the region $[0,a_c]$ 
in such a way that it joins Eq.~(\ref{fcout}) smoothly at 
$\rho=a_c$. This can 
be performed in a basis expansion method 
as explained in Appendix A. In order for this method to work, 
we have to make sure that the strength function is stable 
with respect to the change of $a_c$.

\subsection{Normalization of continuum states}
\label{normalization}
To discuss the normalization of the continuum state in CDA case, 
we first consider the plane wave 
$(2\pi)^{-3}{\rm e}^{i{\vis k}_1\cdot{\vis x}_1+
i{\vis k}_2\cdot{\vis x}_2}$, which is expanded 
as follows~\cite{danilin}
\begin{equation}
(2\pi)^{-3}{\rm e}^{i{\vis k}_1\cdot{\vis x}_1+i{\vis k}_2\cdot{\vis
 x}_2}
=(k\rho)^{-2}\sum_{\ell_1 \ell_2KLM_L}i^KJ_{K+2}(k\rho)
{\cal F}^{\ell_1\ell_2}_{KLM_L}(\Omega)
({\cal F}^{\ell_1\ell_2}_{KLM_L}(\Omega_k))^*,
\end{equation}
where $k^2=k_1^2+k_2^2$ and $\Omega_k$ denotes the five angles
constructed from ${\vi k}_1$ and ${\vi k}_2$ in exactly the same manner 
as $\Omega$. 
Using this expansion and the completeness relation~(\ref{complete}), 
we have 
\begin{equation}
\int_0^{\infty}\rho
 J_{K+2}(k\rho)J_{K+2}(k'\rho)d\rho=\frac{1}{k}\delta(k-k').
\label{JJ.int}
\end{equation}
The general form of the properly normalized 
free-wave that has energy $E$ and the 
angular momentum $L$ and its projection $M_L$ is 
\begin{equation}
\Psi_{kLM_L}^{\rm FW}=C_k 
\sum_{\ell_1 \ell_2 K}C^{\ell_1 \ell_2}_K(k)
(k\rho)^{-2}J_{K+2}(k\rho)
{\cal F}^{\ell_1 \ell_2}_{KLM_L}(\Omega),
\label{3b.pw}
\end{equation}
where the amplitudes $C^{\ell_1 \ell_2}_K(k)$ satisfy 
the condition 
$\sum_{\ell_1 \ell_2 K}|C^{\ell_1 \ell_2}_K(k)|^2=1$ and 
$C_k$ is a normalization constant on the energy scale, 
$\langle \Psi_{k'LM_L}^{\rm FW}|\Psi_{kLM_L}^{\rm FW}\rangle=
\delta(E-E')$. 
Using Eq.~(\ref{JJ.int}) and 
$\delta(E-E')=\delta((\hbar^2/2m)(k^2-k'^2))=
(m/\hbar^2k)\delta(k-k')$, $C_k$ becomes 
\begin{equation}
C_k={\cal J}^{-1}\frac{\sqrt{m}}{\hbar}k^2.
\end{equation}

With reference to the above result, the normalization of $f_c^{\rm GF}$  
may be chosen as 
\begin{eqnarray}
f_{kc}^{\rm GF}(\rho)=k^{-5/2}C_k ||f_c^{\rm CD}||
\left|\frac{\cos \delta_c}{\lambda_c}\right|f_c^{\rm GF}(\rho),
\end{eqnarray}
where  $||f_c^{\rm CD}||=[\int_0^{\infty}(f_c^{\rm CD}(\rho))^2d\rho]^{1/2}$ is 
calculated from Eq.~(\ref{spectr.amp}) using $\Psi^{\rm CD}$ and has 
the property $\sum_c ||f_c^{\rm CD}||^2$=1 provided that $\Psi^{\rm CD}$ 
is normalized. A continuum state $\Psi_{kLM_L}^{\rm GF}$ normalized on the 
energy scale is given by
\begin{equation}
\Psi_{kLM_L}^{\rm GF}=\rho^{-5/2}\sum_c f_{kc}^{\rm GF}(\rho){\cal F}_{cLM_L}.
\end{equation}

It should be stressed that the CDA method of constructing the 
continuum state does not require solving the coupled equations for 
$f_c(\rho)$ but only needs the CDS.
However, it does not allow chosing an arbitrary energy. 
Then the discretized state 
is expanded into the channel components and the tail behavior of each 
hyperradial part is readily corrected with the Green's function.

\section{Specific examples}
\label{sec.example}

The masses of three particles are set equal, $A_1=A_2=A_3=1$, in 
unit of $\hbar^2/m=41.47106$\,MeVfm$^2$. One of the particles has 
charge $e$ and others are neutral. 
The three particles are assumed to interact via a hyperscalar 
potential which depends on the hyperradius only 
\begin{equation}
V=V(\rho).
\end{equation}
As commented below Eq.~(\ref{jacobian}), $\rho$ scales the size of 
the system, and the potential depending on $\rho$ is considered a 
special three-body force. 
No channel coupling occurs for this potential in the HH formalism, and 
the three-body problem actually reduces to an easily solvable 
potential problem. The ground state 
consists of a single channel $c_0=(0,0,0)$, and its wave function 
takes the form 
$\rho^{-5/2}f_{c_0}(\rho){\cal F}^{00}_{000}(\Omega)$. 
The excited states with $L=1$ 
which are excited from the ground state by the electric dipole operator  
have two 
channels, $c_1=(0,1,1)$ and $c_2=(1,0,1)$, and they are degenerate 
in energy. Their wave functions are given by 
$\rho^{-5/2}f_{c_1}(\rho){\cal F}^{01}_{11M_L}(\Omega)$ and 
$\rho^{-5/2}f_{c_2}(\rho){\cal F}^{10}_{11M_L}(\Omega)$, respectively. 
Note that the electric dipole operator~(\ref{E1.op}) acting on the 
ground state leads to
\begin{eqnarray}
\frac{1}{d_1} \zeta_{c_1}(\rho)= \frac{1}{d_2} \zeta_{c_2}(\rho)
= \frac{1}{\sqrt{8\pi}}\rho f_{c_0}(\rho).
\label{WPsi}
\end{eqnarray}
Thus the electric dipole strength function is easily evaluated if the 
hyperradial functions $f_c(\rho)$ are obtained. 

The function $f_c(\rho)$ is determined from Eq.~(\ref{eq.f-z}). Since 
$z_c(\rho)$ of Eq.~(\ref{zzeta}) reduces to $V(\rho)f_c(\rho)$ 
for $V=V(\rho)$, Eq.~(\ref{eq.f-z}) simplifies to 
\begin{equation}
\left(\frac{d^2}{d\rho^2}
-\frac{(K+\frac{3}{2})(K+\frac{5}{2})}{\rho^2}-\frac{2m}{\hbar^2}
V(\rho)+k^2\right)f_{c}(\rho)=\frac{2m}{\hbar^2}\zeta_c(\rho).
\label{hypscalar.eq}
\end{equation}
In the CDA case of $\zeta_c(\rho)=0$ we can solve this equation for 
both bound and continuum states 
with high precision 
using, e.g., the Numerov method. With this solutions 
we can numerically obtain the exact electric dipole strength function, 
and 
assess the various methods described in the previous section by 
comparing their strength functions with the exact one. In the DEM case 
where $\zeta_c(\rho)$ does not vanish, the above equation can easily be 
solved using the Bloch operator formalism~\cite{bloch,hesse98,desc06} 
or the method of Appendix A.

For the CSM to be applicable, the potential $V(\rho)$ 
should satisfy analyticity, and a square well potential or a 
Woods-Saxon potential with a small diffuseness parameter has to be 
avoided. The form factor of $V(\rho)$ is assumed to be Gaussian, 
$V_0\,{\rm exp}(-\kappa \rho^2)$. 
To generate the electric dipole strength functions of different 
shape, we chose four sets: Three of them are one-ranged Gaussian 
and the other is three-ranged. The strength and range parameters 
are listed in Table~\ref{table1}. The ground state energies $E_0$ 
and the root mean square value, $\sqrt{\langle\rho^2\rangle}$, of the 
ground state are also listed in the table.  The Set~4 potential 
combined with the effective centrifugal barrier $35\hbar^2/(8m\rho^2)\, (K=1)$ 
has double minima at about 
$\rho=1.4$ and 5.0\,fm, producing a strength function 
of complex shape. 

\begin{table}[t]
\caption{The potential parameters and 
the ground state properties. Set 4 potential is three-ranged. 
Energy and length are given in units of MeV and fm. }
\begin{center}
\begin{tabular}{ccrcccccccccc}
\hline\hline
Set &&$V_0$&&&&$\kappa$&&&$E_0$&&&$\sqrt{\langle\rho^2\rangle}$\\
\hline
1 && $-$110 &&&& 0.16  &&& $-$17.6 &&& 1.39 \\
2 && $-$90  &&&& 0.16  &&& $-$8.95 &&& 1.57 \\
3 && $-$75  &&&& 0.16  &&& $-$3.49 &&& 1.84 \\ 
4 && $-$610 &&&& 0.25  &&& $-$38.4 &&& 0.938\\
  && 570    &&&& 0.16  &&&         &&&      \\
  && $-$200 &&&& 0.10  &&&         &&&      \\
\hline\hline
\end{tabular}
\end{center}
\label{table1}
\end{table}

We used the functions 
\begin{equation}
\phi(a_i)=\rho^{K+5/2}\exp(-{\textstyle \frac{1}{2}}a_i\rho^2)
\label{gauss.expansion}
\end{equation}
as square-integrable bases for an expansion of 
$f_c(\rho)$. The parameters $a_i$ 
were chosen in a geometric progression, $a_i=1/(b_0p^{i-1})^2$ with 
$b_0=0.1$~fm and $p=1.3$, 
to cover a wide $\rho$-space. The maximum number of terms 
was increased up to $i=31$ to assure the convergence of the calculation, 
particularly in the LIT case. 
The electric dipole strength function shown below is not $S(E)$ but 
$s(E)=S(E)/(d_1^2+d_2^2)$ in unit of ${\rm fm}^2\, {\rm MeV}^{-1}$.

\subsection{Result with CDA}

\begin{figure}
\centerline{\includegraphics[width=9cm]{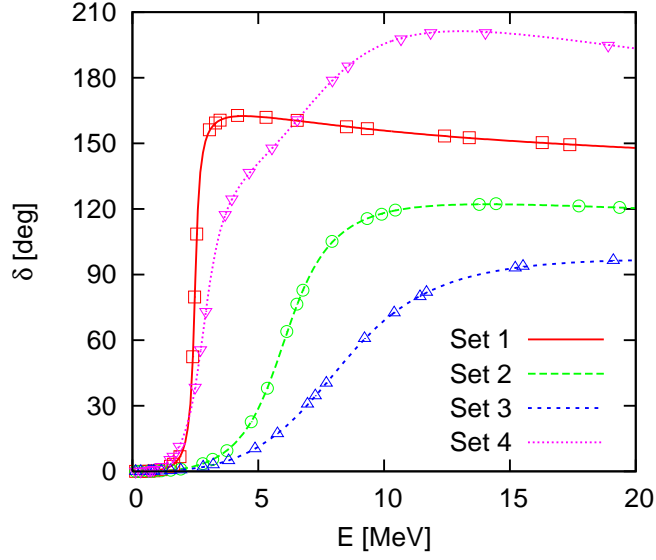}}
\caption{(Color online) Comparison of the $P$-wave $(K=1)$ phase shifts 
calculated with Numerov and CDA methods 
for the potential sets of Table~\ref{table1}.} 
\label{fig1}
\end{figure}

\begin{figure}
\centerline{\includegraphics[width=14cm]{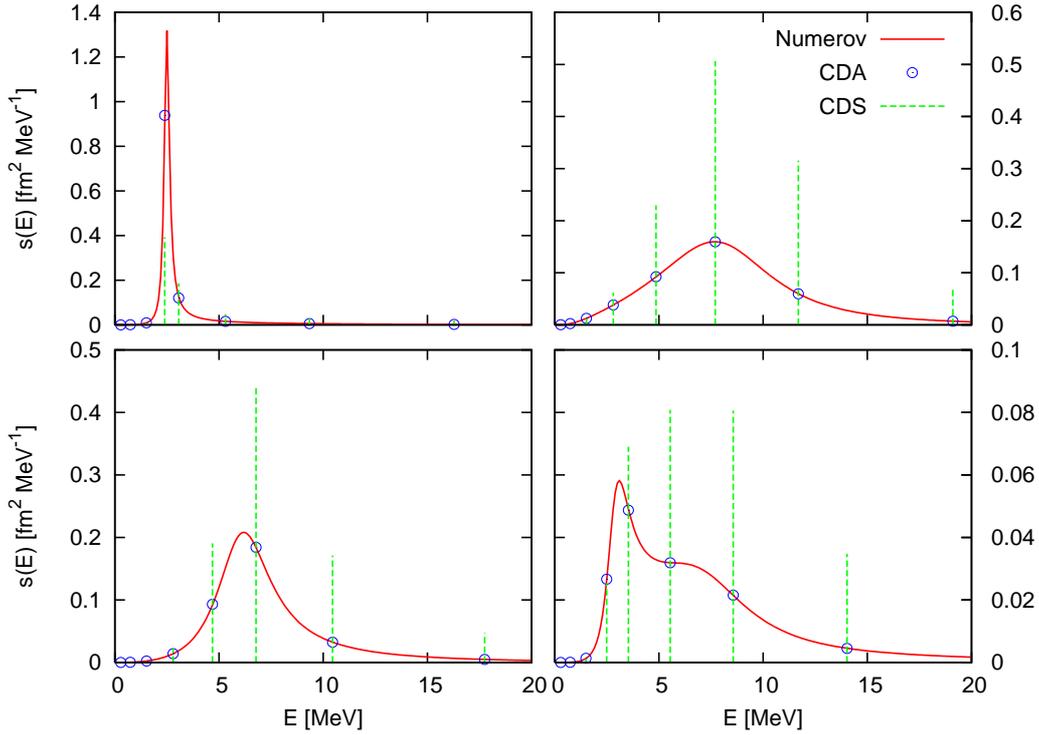}}
\caption{(Color online) Comparison of the electric dipole strength function 
between Numerov and CDA methods. Dotted vertical lines 
stand for the strength calculated with CDS. Potentials: Set 1~(Left
 upper), Set 2~(Left lower), Set 3~(Right upper), Set 4~(Right lower).} 
\label{fig2}
\end{figure}

Equation~(\ref{hypscalar.eq}) with $\zeta_c(\rho)=0$ 
is solved in a combination of basis functions $\phi(a_i)$ 
in order to get the CDS, $\Psi^{\rm CD}$. 
Figure~\ref{fig1} compares the phase shifts for $K=1\, (L=1)$ 
between the Numerov 
and CDA methods. The CDA phase shift shown in symbols is calculated from 
Eq.~(\ref{ps.formula}).  None of the sets 
supports a bound state with $L=1$, but Set 1 potential produces a sharp 
resonance around $E=2.5$\,MeV. 
The resonance is shifted to 
higher energy and becomes broader for the potentials of 
Sets 2 and 3. 
The CDA calculation reproduces the exact phase shifts fairly 
well for all four potentials. This confirms that 
the wave function constructed from the discretized state
with the help of the Green's function can describe the 
continuum accurately. The CDA can thus 
reproduce the strength functions without question, as shown 
in Fig.~\ref{fig2}. Correcting the tail behavior of the 
discretized state is therefore very useful to predict the exact strength. 
The dotted vertical line in Fig.~\ref{fig2} stands for the strength calculated 
from the CDS, which ignores the continuum effect. The CDS strength is 
in reasonable correspondence to that of CDA near the sharp resonance 
region at about $E=2$-3 MeV for Set 1 potential, 
but it generally exhibits a strong deviation at higher energies. 

\subsection{Result with DEM}

\begin{figure}
\centerline{\includegraphics[width=14cm]{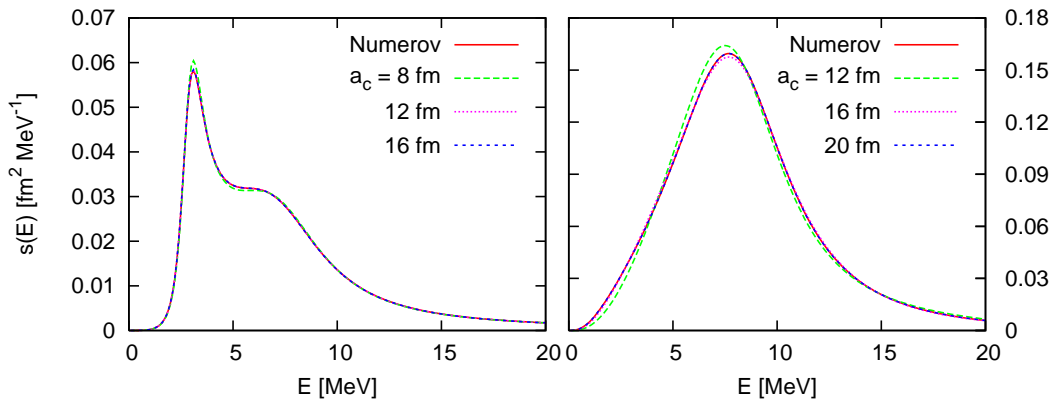}}
\caption{(Color online) Comparison of the electric dipole strength function 
for Set 4~(Left) and Set 2~(Right) potentials between Numerov 
and DEM calculations.} 
\label{fig3}
\end{figure}

To solve Eq.~(\ref{hypscalar.eq}) in the DEM method following Appendix
A, we again use the same basis functions $\phi(a_i)$. 
Figure~\ref{fig3} displays the $s(E)$ values for Set 4 and Set 2  
potentials obtained in the DEM calculation. The strength for Set 4 becomes  
stable for $a_c \gtrsim 10$~fm, while a larger $a_c$ is required 
for the Set 2 case to obtain the stability. 
The performance in DEM is very satisfactory for all the potential sets, that
is, accurate strength functions are obtained independently of the
potentials once $a_c$ is chosen to be large enough.  

\subsection{Result with CSM}

\begin{figure}
\centerline{\includegraphics[width=8cm]{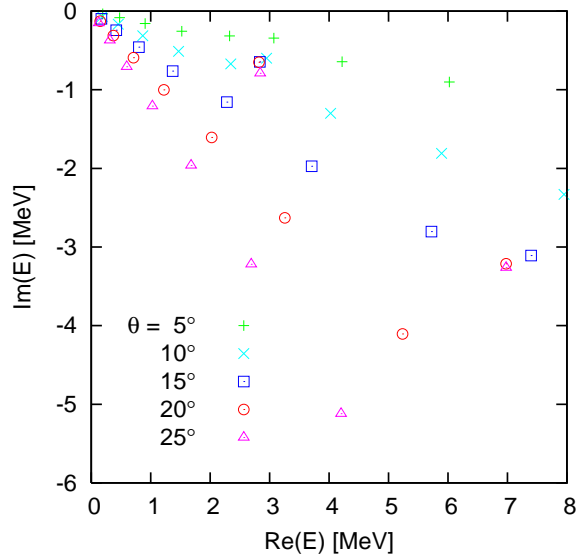}}
\caption{(Color online) Complex eigenvalues with $K=1$, $L=1$ for Set 4 
potential as a function of the scaling angle $\theta$.}
\label{fig4}
\end{figure}

\begin{figure}
\centerline{\includegraphics[width=9cm]{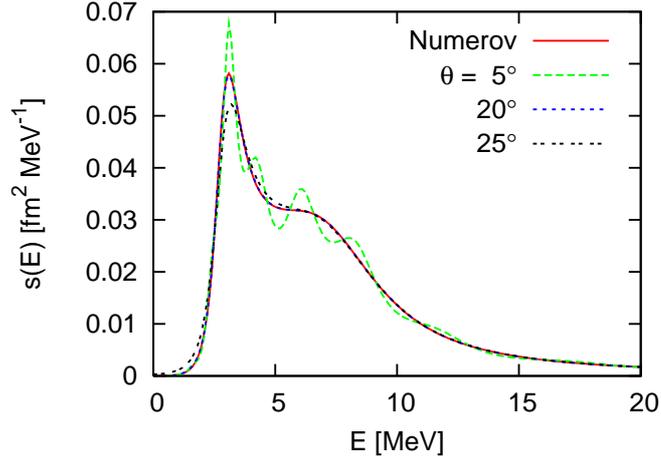}}
\caption{(Color online) Comparison of the electric dipole strength function 
for Set 4 potential between Numerov and CSM calculations. } 
\label{fig5}
\end{figure}

In the CSM
calculation $\rho$ is transformed to ${\rm e}^{i\theta}\rho$. 
The diagonalization of the rotated Hamiltonian 
is performed using the basis functions $\phi(a_i)$. Figure~\ref{fig4} 
displays the complex eigenvalues $E^{\lambda}(\theta)$ for 
Set 4 potential for some $\theta$ values. 
We observe two eigenvalues which are rather stable 
with respect to the change of $\theta$. One of them corresponds to a 
sharp resonance at about $E=$~3 MeV in Fig.~\ref{fig1}, and 
another to a very broad peak around 7 MeV. 
The latter shows up for $\theta \geq 15^{\circ}$. Figure~\ref{fig5} 
displays how the strength function changes as a function of $\theta$. The 
scaling angle $\theta$ is changed up to 25$^\circ$ to see the 
dependence of $s(E)$ on $\theta$. 
The CSM reproduces the exact strength 
very well when $\theta$ is taken in the range 
15$^\circ \lesssim \theta \lesssim 20^\circ$ which covers 
the two stable eigenvalues noted above. 
Increasing $\theta$ beyond 25$^\circ$ begins to deteriorate the 
agreement with the exact strength function.

\begin{figure}
\centerline{\includegraphics[width=14cm]{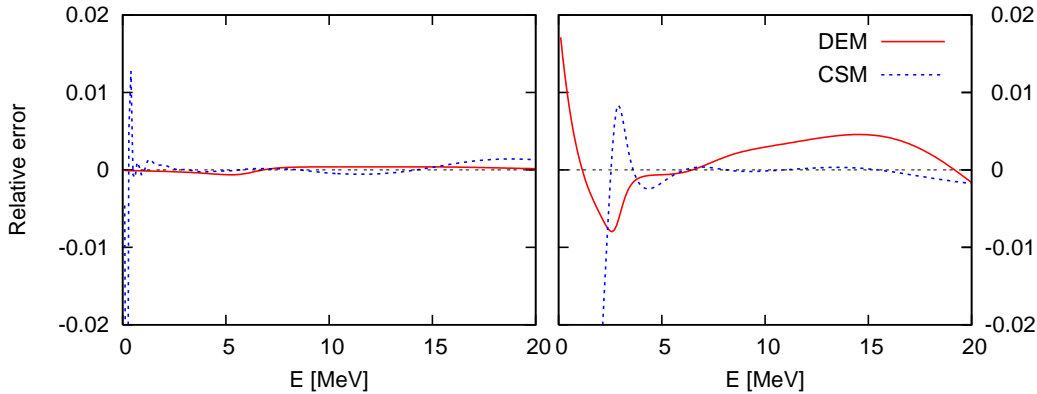}}
\caption{(Color online) Relative error of the electric dipole 
strength function for Set 2~(Left) and Set 4~(Right) potentials.} 
\label{fig6}
\end{figure}

Figure~\ref{fig6} plots the relative error, 
$(s_{\rm exact}(E)-s(E))/s_{\rm exact}(E)$, of the strength 
functions calculated with the CSM and DEM models, where 
$s_{\rm exact}(E)$ stands for the strength function calculated by 
the Numerov method. Both CSM and DEM reproduce the exact strength 
very well (within 1 \%) over the wide energy range except that 
the CSM tends to give large errors near the threshold energy. 
The reason for this is understood from the fact that a correct energy 
dependence of $s(E)$ for $E$ close to zero is not manifestly  
guaranteed in the CSM. The DEM can incorporate such a behavior 
because it 
solves the driven equation of motion directly.

\subsection{Result with LIT}

\begin{figure}
\centerline{\includegraphics[width=9cm]{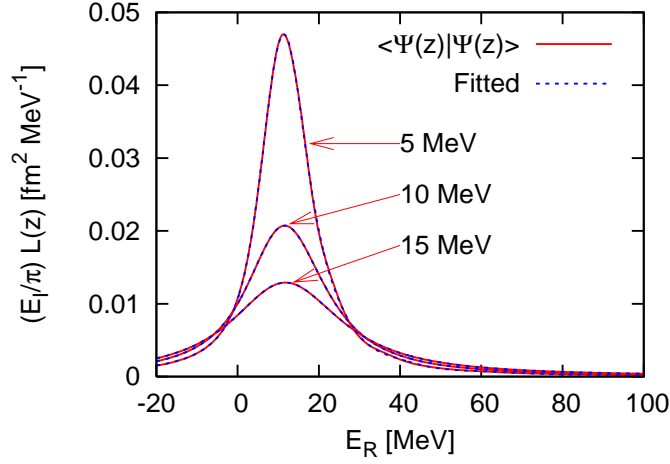}}
\caption{(Color online) Comparison of the Lorentz transform as a function
 of $E_R$ for $E_I=5, 10, 15$~MeV. 
Solid lines are obtained as the norm of $\Psi(z)$, while
 dotted lines are the integral transform of Eq.~(\ref{algebraic}) 
with the Lorentz kernel.   Set~3 potential is employed.} 
\label{fig7}
\end{figure}

The same type of bases as in CDA is used to calculate 
the Lorentz transforms ${\cal L}(z)$. Since we need ${\cal L}(z)$ in a 
wide $E_R$ region, the falloff parameters $a_i$ in 
Eq.~(\ref{gauss.expansion}) are taken in a region wide 
enough to describe various 
shapes of the function $\Psi(z)$. To make the inversion from 
${\cal L}(z)$ to $S(E)$, it is convenient to express $S(E)$ in terms of 
some plausible functions which the dipole strength of three particles is
expected to take. As discussed in Appendix C, $S(E)$ should show 
$E^3$ dependence near $E=0$, so we assume the following form 
\begin{equation}
s(E)=E^3 \sum_{n=1}^N C_n \exp\Big(-\frac{\alpha}{n}E\Big),
\label{algebraic}
\end{equation}
which is often employed in the literature~\cite{efros}. The 
${\cal L}(z)$ values calculated as the norm of $\Psi(z)$ are fitted from 
those calculated from the above $s(E)$ using 
Eq.~(\ref{L-S.relation}). 
The coefficients $C_n$ are determined by the least squares fitting. The 
parameter $\alpha$ and the maximum number $N$ of terms are varied to
obtain ``converging'' $s(E)$ as much as possible. 
Figure~\ref{fig7} compares the two ${\cal L}(z)$ for some $E_I$ values. 

\begin{figure}
\centerline{\includegraphics[width=9cm]{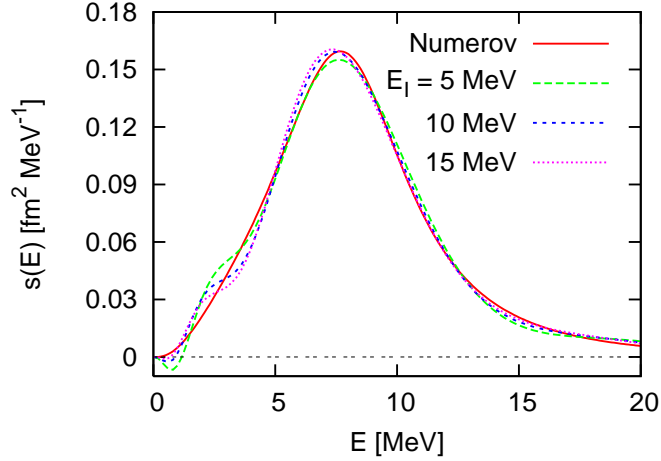}}
\caption{(Color online) Comparison of the electric dipole strength function 
between the Numerov method and the LIT methods calculated 
for $E_I=5, 10, 15$~MeV. 
Set~3 potential is employed.} 
\label{fig8}
\end{figure}

Figure~\ref{fig8} displays the corresponding $s(E)$ functions. Though
the fitting of ${\cal L}(z)$ appears to be very satisfactory except for the 
case of small $E_I$, the resulting $s(E)$ functions differ from each
other, exhibiting some
deviations from the exact strength function in the energy region below 
the peak of the strength. Particularly the strength near the threshold 
shows an oscillatory behavior and becomes even negative. The latter is 
unavoidable in general because the expression~(\ref{algebraic}) does not 
guarantee the positive definiteness. We tested another form 
\begin{equation}
s(E)=E^3 \left(\sum_{n=1}^N C'_n
	  \exp\Big(-\frac{\alpha'}{n}E\Big)\right)^2, 
\end{equation}  
which is always positive. 
In this case determining the coefficients $C'_n$ is not so easy. We used 
the amoeba routine~\cite{recipes} to search for a minimum of a
function. Since it is hard to increase $N$ in this routine, no better 
result was attainable. 

In such a case that the strength function has a very narrow peak or
complex shape as a function of energy, the assumption of $S(E)$ becomes  
in general much less trivial. Though there is some 
effort~\cite{efros,andreasi,leidemann}, 
no convincing form of the strength function or inverting method 
is available.

\section{Conclusion}
\label{sec.conclusion}

The electric dipole strength functions provide interesting possibilities 
of comparisons between theory and experiment. 
They require accurate theoretical calculations which become very difficult 
for a three-body system. 
In recent years, the three-body continuum has started to play a vital role 
in the study of the excitation mechanism of two-neutron halo nuclei. 
In the present paper we have developed and compared practical 
methods to compute dipole strengths for a three-body system with a 
discretized continuum. 
Using discrete square-integrable states has the important advantage 
that powerful techniques developed for bound-state variational studies 
can be employed such as the versatile expansions in Gaussian states. 

The new techniques to compute the strength function with a discrete basis 
are a direct approach with discrete states (CDA method) and the calculation 
of a summed expression of the strength function involving a solution 
of an inhomogeneous driven equation of motion (DEM method). 
They both make use of Green's functions. 
We have compared them with the complex scaling method (CSM) and the Lorentz 
integral transform (LIT), also making use of a discretized continuum. 
An apparent advantage of the CSM and LIT is that strength functions 
can be obtained from states with unphysical asymptotic behaviors, 
i.e.\ asymptotically vanishing functions. 
However, the Green's function technique leads to physical asymptotics. 
In the CDA, the correct asymptotic behavior of scattering states 
is first obtained, allowing a direct calculation of the dipole strength. 
In the DEM, the solution of the driven equation of motion has a three-body 
outgoing-wave behavior. 

Numerical tests have been performed with an hyperscalar three-body 
potential in the hyperspherical-harmonics formalism. 
They have shown that the LIT method is not as mature as the other ones. 
The problems encountered with the LIT are not due to the discretized continuum 
but to the difficulty of accurately inverting the obtained transform. 
Any significant progress in this inversion would improve the present results. 
All other methods have provided comparable accurate results. 
The CSM presents however some accuracy problems at very low energies. 

The present approaches are promising tools for the study of realistic 
three-body strength functions with full account of channel couplings. 
In the three-body continuum, an infinity of open channels occur at each total 
energy. 
They correspond to the various ways of how this energy can be shared between 
the particles. 
The necessary extensions will also have to deal with Coulomb potentials 
for which couplings occur even at large distances.

\section*{Acknowledgments}

We would like to thank S. Aoyama, W. Leidemann and 
W. Vanroose for useful and enlightening discussions. 
This work was supported in part by a Grant-in-Aid for Scientific
Research (No.~21540261) and also by 
the Bilateral Joint Research Project between 
the JSPS (Japan) and the FNRS (Belgium).  
W. H. is supported by a Grant-in Aid for Scientific Research for 
Young Scientists (No.~19$\cdot$3978) as a JSPS Research Fellow 
for Young Scientists. 
This text presents research results of the Belgian program P6/23
on interuniversity attraction poles initiated by the Belgian
Federal Science Policy Office.

\appendix

\section{Solution of an inhomogeneous equation with a boundary
 condition}

The aim of this appendix is 
to solve Eq.~(\ref{y.formal}) together with (\ref{z.exp}), where 
$z_c(\rho)$ and $\zeta_c(\rho)$ are assumed to vanish for 
$\rho \geq a_c$. From Eq.~(\ref{fcout}), 
the logarithmic derivative of $f_c(\rho)$ at $\rho=a_c$ is a known 
constant given by 
\begin{equation}
B_c=\frac{a_cf_c'(a_c)}{f_c(a_c)}=\frac{a_c h_c^{+ \prime}(a_c)}
{h_c^+(a_c)}.
\end{equation}
We thus need to determine $f_c(\rho)$ in the interval 
$[0, a_c]$, which we call $f_c^{\rm int}(\rho)$, with the constraint that 
its logarithmic derivative at $\rho=a_c$ is $B_c$. 

We try to obtain $f_c^{\rm int}(\rho)$ 
in an expansion with some 
basis sets $(\phi_{c1}(\rho),\phi_{c2}(\rho),\ldots,$\\$ \phi_{cn_c}(\rho))$
\begin{equation}
f_c^{\rm int}(\rho)=\sum_{i=1}^{n_c} X_{ci} {\overline \phi}_{ci}(\rho) 
\end{equation}
with 
\begin{equation}
{\overline \phi}_{ci}(\rho)=\phi_{ci}(\rho)-\gamma_{ci} \phi_{c0}(\rho).
\end{equation}
Here $\phi_{c0}(\rho)$ is an auxiliary function introduced to 
satisfy the condition for the logarithmic derivative at $\rho=a_c$. In fact  
the logarithmic derivative of $f_c^{\rm int}(\rho)$ at $\rho=a_c$ 
becomes $B_c$ provided that $\gamma_{ci}$ is chosen as 
\begin{equation}
\gamma_{ci}=\frac{B_c \phi_{ci}(a_c)-a_c \phi'_{ci}(a_c)}{B_c\phi_{c0}(a_c)-a_c \phi'_{c0}(a_c)}.
\end{equation}

The equation to determine the coefficients $X_{ci}$ reads 
\begin{equation}
\sum_{c'}\sum_{j=1}^{n_{c'}}\Big\{({\overline \phi}_{ci}|{\overline \phi}_{c'j})\delta_{c,c'}
-({\overline \phi}_{ci}|{\cal M}_{cc'}|{\overline \phi}_{c'j})
\Big\}X_{c'j}=({\overline \phi}_{ci}|{\cal S}_c),
\end{equation}
where the round brackets indicate that the integration is to be done 
in $[0,a_c]$: 
\begin{eqnarray}
& &({\overline \phi}_{ci}|{\overline \phi}_{c'j})
=\int_0^{a_c} ({\overline \phi}_{ci}(\rho))^*
{\overline \phi}_{c'j}(\rho) d\rho,
\nonumber \\
& &({\overline \phi}_{ci}|{\cal M}_{cc'}|{\overline \phi}_{c'j})
=\frac{2m}{\hbar^2}\int_0^{a_c}\!\! \int_0^{a_{c'}} ({\overline \phi}_{ci}(\rho))^* G_c^+(\rho, \rho')
V_{cc'}(\rho'){\overline \phi}_{c'j}(\rho') d\rho d\rho',
\nonumber \\
& &({\overline \phi}_{ci}|{\cal S}_c)=\frac{2m}{\hbar^2}\int_0^{a_c}\!\! 
\int_0^{a_c} ({\overline \phi}_{ci}(\rho))^* G_c^+(\rho, \rho') 
\zeta_c(\rho')d\rho d\rho'. 
\end{eqnarray}
The amplitude 
$A_c$ is determined from $f_c^{\rm int}(a_c)/h_c^+(a_c)$.

\section{Resolvent in complex scaling method}

The complex rotation $U(\theta)$ in the CSM is defined by
\begin{equation}
U(\theta)\Psi({\vi x})=\exp\left({\frac{3}{2}if\theta}\right)
\Psi({\vi x}{\rm e}^{i\theta}),
\end{equation}
where $f$ is the degree of freedom. 
($f=2$ for a three-body system). 
Using $U(\theta)U^{-1}(\theta)$\\$=1$, it is easy to show 
that $U^{-1}(\theta)=U(-\theta)$. A key point is that 
this complex transformation makes the outgoing wave damp asymptotically. 
For a real potential $V({\vi x})$, the Hermitian conjugation 
of $H(\theta)$ is $H(\theta)^{\dagger}=H(-\theta)=H(\theta)^*$, 
where $^*$ denotes the complex conjugation.
Note that the eigenfunctions of $H(\theta)$, 
$\Psi^{\lambda}(\theta)$ and 
$\Psi^{\lambda'}(\theta)$ of Eq.~(\ref{diag.csm}), 
with different labels 
$\lambda$ and $\lambda'$ are not orthogonal in general. 

Together with Eq.~(\ref{diag.csm}), we consider an accompanying 
eigenvalue problem 
\begin{equation}
{\widetilde H}(\theta){\widetilde \Psi}(\theta)={\widetilde E}(\theta)
{\widetilde \Psi}(\theta),
\label{tildeeq.csm}
\end{equation} 
with ${\widetilde H}(\theta)=H(\theta)^*=H(\theta)^{\dagger}$. 
The solution
of this equation is labeled by the same $\lambda$ as 
that of Eq.~(\ref{diag.csm}), and we may 
choose the solution as follows: 
\begin{equation}
{\widetilde \Psi}^{\lambda}(\theta)=(\Psi^{\lambda}(\theta))^*,\ \ \
 \ \ {\widetilde E}^{\lambda}(\theta)=(E^{\lambda}(\theta))^*.
\end{equation}
We can show that both sets of $\{\Psi^{\lambda}(\theta)\}$ and 
$\{{\widetilde \Psi}^{\lambda}(\theta)\}$ are 
biorthogonal~\cite{berggren}, that is, 
\begin{equation}
\langle {\widetilde \Psi}^{\lambda'}(\theta)|
 \Psi^{\lambda}(\theta)\rangle =\delta_{\lambda, \lambda'},
\label{normalization.CSM}
\end{equation}
if the normalization of $\Psi^{\lambda}(\theta)$ is chosen to 
satisfy 
\begin{equation}
\int (\Psi^{\lambda}(\theta))^2 d{\vi x}=1.
\label{norm}
\end{equation}
To prove Eq.~(\ref{normalization.CSM}), we start from 
\begin{equation}
\langle {\widetilde \Psi}^{\lambda'}(\theta)|H(\theta)|
 \Psi^{\lambda}(\theta)\rangle=E^{\lambda}(\theta)
\langle {\widetilde \Psi}^{\lambda'}(\theta)|
 \Psi^{\lambda}(\theta)\rangle.
\end{equation}
The left-hand side is reduced to 
\begin{equation}
\langle {\widetilde \Psi}^{\lambda'}(\theta)|H(\theta)|
 \Psi^{\lambda}(\theta)\rangle=
\langle {\widetilde H}(\theta){\widetilde \Psi}^{\lambda'}(\theta)|
 \Psi^{\lambda}(\theta)\rangle=
E^{\lambda'}(\theta)\langle {\widetilde \Psi}^{\lambda'}(\theta)|
 \Psi^{\lambda}(\theta)\rangle.
\end{equation}
Thus for $E^{\lambda}(\theta)\neq E^{\lambda'}(\theta)$, 
$\langle {\widetilde \Psi}^{\lambda'}(\theta)|
 \Psi^{\lambda}(\theta)\rangle=0$, which, together with 
Eq.~(\ref{norm}), leads to the biorthogonality 
relation~(\ref{normalization.CSM}).

It follows from the biorthogonality that the resolvent 
can be expanded as
\begin{equation}
R(\theta)=\sum_{\lambda}\frac{1}{E-E^{\lambda}(\theta)+i\epsilon}
|\Psi^{\lambda}(\theta)\rangle 
\langle {\widetilde \Psi}^{\lambda}(\theta)|.
\end{equation}
Substitution of this $R(\theta)$ into Eq.~(\ref{dbdecsm}) 
leads to the strength function~(\ref{str.csm}), but 
${\cal D}^{\lambda}(\theta)$ and 
$\widetilde{{\cal D}}^{\lambda}(\theta)$ are defined by 
\begin{equation}
{\cal D}^{\lambda}(\theta)=
\langle {\widetilde \Psi}^{\lambda}(\theta)|U(\theta)W\Psi_0 \rangle,\ \ \ \ \ 
\widetilde{{\cal D}}^{\lambda}(\theta)=
\langle \Psi_0 | W^{\dagger} U^{-1}(\theta)\Psi^{\lambda}(\theta)\rangle.
\end{equation}
These are shown to be identical to those given in 
Eq.~(\ref{def.DD}). The equality for ${\cal D}^{\lambda}(\theta)$ 
is trivial from the definition of $W(\theta)$. For the case of 
$\widetilde{{\cal D}}^{\lambda}(\theta)$, 
we only need to show that $\langle \Psi_0 | U^{-1}(\theta)$ and  
$\langle (U(\theta)\Psi_0)^* |$ are identical within a 
$\theta$-independent phase which may be chosen to be unity.
This is justified by comparing 
the normalization condition, 
$\langle (U(\theta)\Psi_0)^* | U(\theta)\Psi_0 \rangle=1$, 
with the trivial normalization condition for $\Psi_0$, 
$\langle \Psi_0 | U^{-1}(\theta)U(\theta)| \Psi_0 \rangle=1$.

\section{Analytic form of electric dipole strength function in three-body 
continuum}

The form of $S(E)$ as a function of $E$ is vital in inverting
Eq.~(\ref{L-S.relation}) which relates $S(E)$ to ${\cal L}(z)$. 
Its form may be discussed by examining the matrix
element $\langle \Psi_{\nu}|W_{\mu}|\Psi_0 \rangle$. 
To this end we assume that 
the continuum state $\Psi_{\nu}$ is approximated by the
free wave~(\ref{3b.pw}). First we consider a case where $E$ is close to
zero.  Since $K=1$ in the present case, the free wave has $k^3$-dependence
for small $k$, that is, the matrix element 
behaves like $k^3\propto E^{3/2}$. Hence $S(E)$ has an $E^3$ dependence 
near the threshold. 
Note that this dependence is different 
from that of a two-body continuum, where the radial part of the 
free wave normalized on
the energy scale is given for the partial wave $\ell$ by 
\begin{equation}
\sqrt{\frac{2\mu k}{\pi \hbar^2}}j_{\ell}(kr).
\end{equation}
Here $\mu$ is the reduced mass of the two particles. 
Thus the electric dipole matrix element for small $k$ scales as 
$k^{3/2}\propto E^{3/4}$ for the $P$-wave.

To know a more general form of $S(E)$, let us assume that the hyperradial 
part $f_{c_0}(\rho)$ of the ground state $\Psi_0$ 
is approximated in terms of a combination of Gaussians, 
$\rho^{5/2}\exp(-{\textstyle \frac{1}{2}}a\rho^2)$, or 
Exponentials,  $\rho^{5/2}\exp(- b\rho)$. 
The $k$-dependence of the matrix element 
becomes $k^3 \exp(-k^2/2a)$ for the Gaussians, while it 
is given by $k^3 (k^2/b^2+1)^{-9/2}$ for the Exponentials. 
The strength close to $E=0$ is
included in these functional forms. The analytic form of $S(E)$ 
we are looking for is therefore 
\begin{equation}
S(E)=E^3 \left(\sum_n C_n {\rm e}^{-\alpha_n E}\right)^2,
\label{gauss.S}
\end{equation}
with $\alpha_n=m/(\hbar^2 a_n)$ for the Gaussian case, or 
\begin{equation}
S(E)=E^3 \left(\sum_n C_n \left(\beta_n E+1\right)^{-9/2}\right)^2,
\label{exp.S}
\end{equation}
with $\beta_n=2m/(\hbar^2 b_n^2)$ for the Exponential case. 

Because the actual continuum deviates from the free wave through the 
interaction among the particles, neither 
Eq.~(\ref{gauss.S}) nor Eq.~(\ref{exp.S}) is exact. However, a suitable 
choice of the coefficients $C_n$ may simulate $S(E)$ accurately.  
The Lorentz transform is calculated from Eq.~(\ref{gauss.S}) or
Eq.~(\ref{exp.S}) according to Eq.~(\ref{L-S.relation}). 
The coefficients $C_n$ 
are then determined by least squares fitting to the overlap function 
${\cal L}(z)$, which is in general never trivial particularly when 
the number of terms is great. It is thus more popular to use non-squared 
from as in Eq.~(\ref{algebraic}).


\begin{thebibliography}{99}
\bibitem{danilin}B.V. Danilin, I.J. Thompson, J.S. Vaagen, and
	M.V. Zhukov, Nucl. Phys. A\,{\bf 632} (1998), 383. 
\bibitem{danilin06}B.V. Danilin, J.S. Vaagen, T. Rogde, S.N. Ershov, 
I.J. Thompson, and M.V. Zhukov, Phys. Rev. C\,{\bf 73} (2006), 054002.
\bibitem{garrido}E. Garrido, A.S. Jensen, and D.V. Fedorov,
	Phys. Rev. C\,{\bf 78} (2008), 034004.
\bibitem{bk}P. Barletta and A. Kievsky, Few-Body Syst. {\bf 45} (2009) 25.
\bibitem{barletta} P. Barletta, C. Romeo-Redondo, A. Kievsky, 
M. Viviani, and E. Garrido, Phys. Rev. Lett.\,{\bf 103} (2009), 090402.
\bibitem{nielsen}E. Nielsen, D.V. Fedorov, A.S. Jensen, and 
E. Garrido,	Phys. Rep.\,{\bf 347} (2001), 373.
\bibitem{desc06}P. Descouvemont, E. Tursunov, and D. Baye, Nucl. 
Phys. A\,{\bf 765} (2006), 370.
\bibitem{aoyama}S. Aoyama, T. Myo, K. Kat\={o} and K. Ikeda,
	Prog. Theor. Phys.\,{\bf 116} (2006), 1.
\bibitem{efros}V.D. Efros, W. Leidemann, G. Orlandini, and N. Barnea,
	J. Phys. G: Nucl. Part. Phys.\,{\bf 34} (2007), R459. 
\bibitem{baye}D. Baye, P. Capel, P. Descouvemont, and Y. Suzuki,
	Phys. Rev. C\,{\bf 79} (2009), 024607.
\bibitem{myo}T. Myo, K. Kat\={o}, S. Aoyama, and K. Ikeda,
	Phys. Rev. C\,{\bf 63} (2001), 054313.
\bibitem{egami}T. Egami, T. Matsumoto, K. Ogata, and M. Yahiro,
	Prog. Theor. Phys.\,{\bf 121} (2009), 789.
\bibitem{ps.cal}Y. Suzuki, W. Horiuchi, and K. Arai, 
Nucl. Phys. A\,{\bf 823}, (2009), 1.
\bibitem{shlomo}S. Shlomo and G. Bertsch, Nucl. Phys. A\,{\bf 243} (1975), 507.
\bibitem{matsuo}M. Matsuo, Nucl. Phys. A\,{\bf 696} (2001), 371.
\bibitem{khan}E. Khan, N. Sandulescu, M. Grasso, and N. Van Giai,
	Phys. Rev. C\,{\bf 66} (2002), 024309. 
\bibitem{pont} M. Pont and R. Shakeshaft, Phys. Rev. A\,{\bf 51} (1995),
	494.
\bibitem{vanroose}W. Vanroose, D.A. Horner, F. Mart\'{i}n, T.N. Rescigno, 
and C.W. McCurdy, Phys. Rev. A\,{\bf 74} (2006), 052702. 
\bibitem{mcmurdy}C.W. McCurdy and F. Mart\'{i}n, J. Phys. B: At. Mol. 
Opt. Phys.\,{\bf 37} (2004), 917.
\bibitem{lowdin}P.O. L\"{o}wdin, in {\it Advances in Quantum Chemistry}
	(Academic Press) {\bf 19} (1988), 87.
\bibitem{kruppa} A.T. Kruppa, R.G. Lovas, and B. Gyarmati,
	Phys. Rev. C\,{\bf 37} (1988), 383.
\bibitem{thompson}I.J. Thompson, B.V. Danilin, V.D. Efros, J.S. Vaagen,
	J.M. Bang, and M.V. Zhukov, Phys. Rev. C\,{\bf 61} (2000), 024318.
\bibitem{bloch} C.~Bloch, Nucl. Phys. {\bf 4} (1957), 503.
\bibitem{hesse98} M. Hesse,  J.-M. Sparenberg,  F. Van Raemdonck,  D. Baye,  
Nucl. Phys. A {\bf 640} (1998), 37.
\bibitem{recipes}W.H. Press, S.A. Teukolsky, W.T. Vetterling, and
	B.P. Flannery, in {\it Numerical recipes in Fortran 77}, (Cambridge
	University Press, New York, 1992).
\bibitem{andreasi}D. Andreasi, W. Leidemann, C. Rei\ss, and M. Schwamb,
	Eur. Phys. J. A. {\bf 24} (2005), 361.
\bibitem{leidemann}W. Leidemann, Few Body Syst. {\bf 42} (2008), 139.
\bibitem{berggren}T. Berggren, Nucl. Phys. A {\bf 109} (1968), 265.

  


\end{thebibliography}
\end{document}